\documentclass[conference]{IEEEtran}

\usepackage{cite}
\usepackage[T1]{fontenc}
\usepackage{graphicx}
\usepackage{amssymb}
\usepackage{amsmath}
\usepackage{amsthm}

\usepackage{paralist}
\usepackage{setspace}
\usepackage{microtype}
\usepackage{url}
\usepackage{balance}
\usepackage{subfigure}
\usepackage{xcolor}
\usepackage{fancyref}
\usepackage{framed}
\usepackage{booktabs}
\usepackage{nicefrac}
\usepackage{multirow}

\usepackage{algorithm}
\usepackage{algpseudocode}

\usepackage{amssymb}
\usepackage{amsfonts}
\usepackage{mathrsfs}
\usepackage{xspace}
\usepackage{bm}
\usepackage{upgreek}

\newcommand{\safemath}[2]{\newcommand{#1}{\ensuremath{#2}\xspace}}

\safemath{\bma}{\mathbf{a}}
\safemath{\bmb}{\mathbf{b}}
\safemath{\bmc}{\mathbf{c}}
\safemath{\bmd}{\mathbf{d}}
\safemath{\bme}{\mathbf{e}}
\safemath{\bmf}{\mathbf{f}}
\safemath{\bmg}{\mathbf{g}}
\safemath{\bmh}{\mathbf{h}}
\safemath{\bmi}{\mathbf{i}}
\safemath{\bmj}{\mathbf{j}}
\safemath{\bmk}{\mathbf{k}}
\safemath{\bml}{\mathbf{l}}
\safemath{\bmm}{\mathbf{m}}
\safemath{\bmn}{\mathbf{n}}
\safemath{\bmo}{\mathbf{o}}
\safemath{\bmp}{\mathbf{p}}
\safemath{\bmq}{\mathbf{q}}
\safemath{\bmr}{\mathbf{r}}
\safemath{\bms}{\mathbf{s}}
\safemath{\bmt}{\mathbf{t}}
\safemath{\bmu}{\mathbf{u}}
\safemath{\bmv}{\mathbf{v}}
\safemath{\bmw}{\mathbf{w}}
\safemath{\bmx}{\mathbf{x}}
\safemath{\bmy}{\mathbf{y}}
\safemath{\bmz}{\mathbf{z}}
\safemath{\bmzero}{\mathbf{0}}
\safemath{\bmone}{\mathbf{1}}

\bmdefine{\biad}{a}
\bmdefine{\bibd}{b}
\bmdefine{\bicd}{c}
\bmdefine{\bidd}{d}
\bmdefine{\bied}{e}
\bmdefine{\bifd}{f}
\bmdefine{\bigd}{g}
\bmdefine{\bihd}{h}
\bmdefine{\biid}{i}
\bmdefine{\bijd}{j}
\bmdefine{\bikd}{k}
\bmdefine{\bild}{l}
\bmdefine{\bimd}{m}
\bmdefine{\bind}{n}
\bmdefine{\biod}{o}
\bmdefine{\bipd}{p}
\bmdefine{\biqd}{q}
\bmdefine{\bird}{r}
\bmdefine{\bisd}{s}
\bmdefine{\bitd}{t}
\bmdefine{\biud}{u}
\bmdefine{\bivd}{v}
\bmdefine{\biwd}{w}
\bmdefine{\bixd}{x}
\bmdefine{\biyd}{y}
\bmdefine{\bizd}{z}

\bmdefine{\bixid}{\xi}
\bmdefine{\bilambdad}{\lambda}
\bmdefine{\bimud}{\mu}
\bmdefine{\bithetad}{\theta}
\bmdefine{\biphid}{\phi}
\bmdefine{\bideltad}{\delta}

\safemath{\bmia}{\biad}
\safemath{\bmib}{\bibd}
\safemath{\bmic}{\bicd}
\safemath{\bmid}{\bidd}
\safemath{\bmie}{\bied}
\safemath{\bmif}{\bifd}
\safemath{\bmig}{\bigd}
\safemath{\bmih}{\bihd}
\safemath{\bmii}{\biid}
\safemath{\bmij}{\bijd}
\safemath{\bmik}{\bikd}
\safemath{\bmil}{\bild}
\safemath{\bmim}{\bimd}
\safemath{\bmin}{\bind}
\safemath{\bmio}{\biod}
\safemath{\bmip}{\bipd}
\safemath{\bmiq}{\biqd}
\safemath{\bmir}{\bird}
\safemath{\bmis}{\bisd}
\safemath{\bmit}{\bitd}
\safemath{\bmiu}{\biud}
\safemath{\bmiv}{\bivd}
\safemath{\bmiw}{\biwd}
\safemath{\bmix}{\bixd}
\safemath{\bmiy}{\biyd}
\safemath{\bmiz}{\bizd}

\safemath{\bmxi}{\bixid}
\safemath{\bmlambda}{\bilambdad}
\safemath{\bmmu}{\bimud}
\safemath{\bmtheta}{\bithetad}
\safemath{\bmphi}{\biphid}
\safemath{\bmdelta}{\bideltad}

\safemath{\bA}{\mathbf{A}}
\safemath{\bB}{\mathbf{B}}
\safemath{\bC}{\mathbf{C}}
\safemath{\bD}{\mathbf{D}}
\safemath{\bE}{\mathbf{E}}
\safemath{\bF}{\mathbf{F}}
\safemath{\bG}{\mathbf{G}}
\safemath{\bH}{\mathbf{H}}
\safemath{\bI}{\mathbf{I}}
\safemath{\bJ}{\mathbf{J}}
\safemath{\bK}{\mathbf{K}}
\safemath{\bL}{\mathbf{L}}
\safemath{\bM}{\mathbf{M}}
\safemath{\bN}{\mathbf{N}}
\safemath{\bO}{\mathbf{O}}
\safemath{\bP}{\mathbf{P}}
\safemath{\bQ}{\mathbf{Q}}
\safemath{\bR}{\mathbf{R}}
\safemath{\bS}{\mathbf{S}}
\safemath{\bT}{\mathbf{T}}
\safemath{\bU}{\mathbf{U}}
\safemath{\bV}{\mathbf{V}}
\safemath{\bW}{\mathbf{W}}
\safemath{\bX}{\mathbf{X}}
\safemath{\bY}{\mathbf{Y}}
\safemath{\bZ}{\mathbf{Z}}

\safemath{\bZero}{\mathbf{0}}
\safemath{\bOne}{\mathbf{1}}
\safemath{\bDelta}{\mathbf{\Delta}}
\safemath{\bLambda}{\mathbf{\UpLambda}}
\safemath{\bPhi}{\mathbf{\Upphi}}
\safemath{\bSigma}{\mathbf{\Upsigma}}
\safemath{\bOmega}{\mathbf{\Upomega}}
\safemath{\bTheta}{\mathbf{\Uptheta}}

\bmdefine{\biAd}{A}
\bmdefine{\biBd}{B}
\bmdefine{\biCd}{C}
\bmdefine{\biDd}{D}
\bmdefine{\biEd}{E}
\bmdefine{\biFd}{F}
\bmdefine{\biGd}{G}
\bmdefine{\biHd}{H}
\bmdefine{\biId}{I}
\bmdefine{\biJd}{J}
\bmdefine{\biKd}{K}
\bmdefine{\biLd}{L}
\bmdefine{\biMd}{M}
\bmdefine{\biOd}{N}
\bmdefine{\biPd}{O}
\bmdefine{\biQd}{P}
\bmdefine{\biRd}{R}
\bmdefine{\biSd}{S}
\bmdefine{\biTd}{T}
\bmdefine{\biUd}{U}
\bmdefine{\biVd}{V}
\bmdefine{\biWd}{W}
\bmdefine{\biXd}{X}
\bmdefine{\biYd}{Y}
\bmdefine{\biZd}{Z}

\bmdefine{\biDelta}{\Delta}
\bmdefine{\biLambda}{\Lambda}
\bmdefine{\biPhi}{\Phi}
\bmdefine{\biSigma}{\Sigma}
\bmdefine{\biOmega}{\Omega}
\bmdefine{\biTheta}{\Theta}

\safemath{\bimA}{\biAd}
\safemath{\bimB}{\biBd}
\safemath{\bimC}{\biCd}
\safemath{\bimD}{\biDd}
\safemath{\bimE}{\biEd}
\safemath{\bimF}{\biFd}
\safemath{\bimG}{\biGd}
\safemath{\bimH}{\biHd}
\safemath{\bimI}{\biId}
\safemath{\bimJ}{\biJd}
\safemath{\bimK}{\biKd}
\safemath{\bimL}{\biLd}
\safemath{\bimM}{\biMd}
\safemath{\bimN}{\biNd}
\safemath{\bimO}{\biOd}
\safemath{\bimP}{\biPd}
\safemath{\bimQ}{\biQd}
\safemath{\bimR}{\biRd}
\safemath{\bimS}{\biSd}
\safemath{\bimT}{\biTd}
\safemath{\bimU}{\biUd}
\safemath{\bimV}{\biVd}
\safemath{\bimW}{\biWd}
\safemath{\bimX}{\biXd}
\safemath{\bimY}{\biYd}
\safemath{\bimZ}{\biZd}

\safemath{\bimDelta}{\biDelta}
\safemath{\bimLambda}{\biLambda}
\safemath{\bimPhi}{\biPhi}
\safemath{\bimSigma}{\biSigma}
\safemath{\bimOmega}{\biOmega}
\safemath{\bimTheta}{\biTheta}

\safemath{\setA}{\mathcal{A}}
\safemath{\setB}{\mathcal{B}}
\safemath{\setC}{\mathcal{C}}
\safemath{\setD}{\mathcal{D}}
\safemath{\setE}{\mathcal{E}}
\safemath{\setF}{\mathcal{F}}
\safemath{\setG}{\mathcal{G}}
\safemath{\setH}{\mathcal{H}}
\safemath{\setI}{\mathcal{I}}
\safemath{\setJ}{\mathcal{J}}
\safemath{\setK}{\mathcal{K}}
\safemath{\setL}{\mathcal{L}}
\safemath{\setM}{\mathcal{M}}
\safemath{\setN}{\mathcal{N}}
\safemath{\setO}{\mathcal{O}}
\safemath{\setP}{\mathcal{P}}
\safemath{\setQ}{\mathcal{Q}}
\safemath{\setR}{\mathcal{R}}
\safemath{\setS}{\mathcal{S}}
\safemath{\setT}{\mathcal{T}}
\safemath{\setU}{\mathcal{U}}
\safemath{\setV}{\mathcal{V}}
\safemath{\setW}{\mathcal{W}}
\safemath{\setX}{\mathcal{X}}
\safemath{\setY}{\mathcal{Y}}
\safemath{\setZ}{\mathcal{Z}}
\safemath{\emptySet}{\varnothing}

\safemath{\colA}{\mathscr{A}}
\safemath{\colB}{\mathscr{B}}
\safemath{\colC}{\mathscr{C}}
\safemath{\colD}{\mathscr{D}}
\safemath{\colE}{\mathscr{E}}
\safemath{\colF}{\mathscr{F}}
\safemath{\colG}{\mathscr{G}}
\safemath{\colH}{\mathscr{H}}
\safemath{\colI}{\mathscr{I}}
\safemath{\colJ}{\mathscr{J}}
\safemath{\colK}{\mathscr{K}}
\safemath{\colL}{\mathscr{L}}
\safemath{\colM}{\mathscr{M}}
\safemath{\colN}{\mathscr{N}}
\safemath{\colO}{\mathscr{O}}
\safemath{\colP}{\mathscr{P}}
\safemath{\colQ}{\mathscr{Q}}
\safemath{\colR}{\mathscr{R}}
\safemath{\colS}{\mathscr{S}}
\safemath{\colT}{\mathscr{T}}
\safemath{\colU}{\mathscr{U}}
\safemath{\colV}{\mathscr{V}}
\safemath{\colW}{\mathscr{W}}
\safemath{\colX}{\mathscr{X}}
\safemath{\colY}{\mathscr{Y}}
\safemath{\colZ}{\mathscr{Z}}

\safemath{\opA}{\mathbb{A}}
\safemath{\opB}{\mathbb{B}}
\safemath{\opC}{\mathbb{C}}
\safemath{\opD}{\mathbb{D}}
\safemath{\opE}{\mathbb{E}}
\safemath{\opF}{\mathbb{F}}
\safemath{\opG}{\mathbb{G}}
\safemath{\opH}{\mathbb{H}}
\safemath{\opI}{\mathbb{I}}
\safemath{\opJ}{\mathbb{J}}
\safemath{\opK}{\mathbb{K}}
\safemath{\opL}{\mathbb{L}}
\safemath{\opM}{\mathbb{M}}
\safemath{\opN}{\mathbb{N}}
\safemath{\opO}{\mathbb{O}}
\safemath{\opP}{\mathbb{P}}
\safemath{\opQ}{\mathbb{Q}}
\safemath{\opR}{\mathbb{R}}
\safemath{\opS}{\mathbb{S}}
\safemath{\opT}{\mathbb{T}}
\safemath{\opU}{\mathbb{U}}
\safemath{\opV}{\mathbb{V}}
\safemath{\opW}{\mathbb{W}}
\safemath{\opX}{\mathbb{X}}
\safemath{\opY}{\mathbb{Y}}
\safemath{\opZ}{\mathbb{Z}}
\safemath{\opZero}{\mathbb{O}}
\safemath{\identityop}{\opI}

\safemath{\veca}{\bma}
\safemath{\vecb}{\bmb}
\safemath{\vecc}{\bmc}
\safemath{\vecd}{\bmd}
\safemath{\vece}{\bme}
\safemath{\vecf}{\bmf}
\safemath{\vecg}{\bmg}
\safemath{\vech}{\bmh}
\safemath{\veci}{\bmi}
\safemath{\vecj}{\bmj}
\safemath{\veck}{\bmk}
\safemath{\vecl}{\bml}
\safemath{\vecm}{\bmm}
\safemath{\vecn}{\bmn}
\safemath{\veco}{\bmo}
\safemath{\vecp}{\bmp}
\safemath{\vecq}{\bmq}
\safemath{\vecr}{\bmr}
\safemath{\vecs}{\bms}
\safemath{\vect}{\bmt}
\safemath{\vecu}{\bmu}
\safemath{\vecv}{\bmv}
\safemath{\vecw}{\bmw}
\safemath{\vecx}{\bmx}
\safemath{\vecy}{\bmy}
\safemath{\vecz}{\bmz}

\safemath{\veczero}{\bmzero}
\safemath{\vecone}{\bmone}
\safemath{\vecxi}{\bmxi}
\safemath{\veclambda}{\bmlambda}
\safemath{\vecmu}{\bmmu}
\safemath{\vectheta}{\bmtheta}
\safemath{\vecphi}{\bmphi}
\safemath{\vecdelta}{\bmdelta}

\safemath{\matA}{\bA}
\safemath{\matB}{\bB}
\safemath{\matC}{\bC}
\safemath{\matD}{\bD}
\safemath{\matE}{\bE}
\safemath{\matF}{\bF}
\safemath{\matG}{\bG}
\safemath{\matH}{\bH}
\safemath{\matI}{\bI}
\safemath{\matJ}{\bJ}
\safemath{\matK}{\bK}
\safemath{\matL}{\bL}
\safemath{\matM}{\bM}
\safemath{\matN}{\bN}
\safemath{\matO}{\bO}
\safemath{\matP}{\bP}
\safemath{\matQ}{\bQ}
\safemath{\matR}{\bR}
\safemath{\matS}{\bS}
\safemath{\matT}{\bT}
\safemath{\matU}{\bU}
\safemath{\matV}{\bV}
\safemath{\matW}{\bW}
\safemath{\matX}{\bX}
\safemath{\matY}{\bY}
\safemath{\matZ}{\bZ}
\safemath{\matzero}{\bmzero}

\safemath{\matDelta}{\bDelta}
\safemath{\matLambda}{\bLambda}
\safemath{\matPhi}{\bPhi}
\safemath{\matSigma}{\bSigma}
\safemath{\matOmega}{\bOmega}
\safemath{\matTheta}{\bTheta}

\safemath{\matidentity}{\matI}
\safemath{\matone}{\matO}

\safemath{\rnda}{A}
\safemath{\rndb}{B}
\safemath{\rndc}{C}
\safemath{\rndd}{D}
\safemath{\rnde}{E}
\safemath{\rndf}{F}
\safemath{\rndg}{G}
\safemath{\rndh}{H}
\safemath{\rndi}{I}
\safemath{\rndj}{J}
\safemath{\rndk}{K}
\safemath{\rndl}{L}
\safemath{\rndm}{M}
\safemath{\rndn}{N}
\safemath{\rndo}{O}
\safemath{\rndp}{P}
\safemath{\rndq}{Q}
\safemath{\rndr}{R}
\safemath{\rnds}{S}
\safemath{\rndt}{T}
\safemath{\rndu}{U}
\safemath{\rndv}{V}
\safemath{\rndw}{W}
\safemath{\rndx}{X}
\safemath{\rndy}{Y}
\safemath{\rndz}{Z}

\safemath{\rveca}{\bimA}
\safemath{\rvecb}{\bimB}
\safemath{\rvecc}{\bimC}
\safemath{\rvecd}{\bimD}
\safemath{\rvece}{\bimE}
\safemath{\rvecf}{\bimF}
\safemath{\rvecg}{\bimG}
\safemath{\rvech}{\bimH}
\safemath{\rveci}{\bimI}
\safemath{\rvecj}{\bimJ}
\safemath{\rveck}{\bimK}
\safemath{\rvecl}{\bimL}
\safemath{\rvecm}{\bimM}
\safemath{\rvecn}{\bimN}
\safemath{\rveco}{\bomO}
\safemath{\rvecp}{\bimP}
\safemath{\rvecq}{\bimQ}
\safemath{\rvecr}{\bimR}
\safemath{\rvecs}{\bimS}
\safemath{\rvect}{\bimT}
\safemath{\rvecu}{\bimU}
\safemath{\rvecv}{\bimV}
\safemath{\rvecw}{\bimW}
\safemath{\rvecx}{\bimX}
\safemath{\rvecy}{\bimY}
\safemath{\rvecz}{\bimZ}

\safemath{\rvecxi}{\bmxi}
\safemath{\rveclambda}{\bmlambda}
\safemath{\rvecmu}{\bmmu}
\safemath{\rvectheta}{\bmtheta}
\safemath{\rvecphi}{\bmphi}

\safemath{\rmatA}{\bimA}
\safemath{\rmatB}{\bimB}
\safemath{\rmatC}{\bimC}
\safemath{\rmatD}{\bimD}
\safemath{\rmatE}{\bimE}
\safemath{\rmatF}{\bimF}
\safemath{\rmatG}{\bimG}
\safemath{\rmatH}{\bimH}
\safemath{\rmatI}{\bimI}
\safemath{\rmatJ}{\bimJ}
\safemath{\rmatK}{\bimK}
\safemath{\rmatL}{\bimL}
\safemath{\rmatM}{\bimM}
\safemath{\rmatN}{\bimN}
\safemath{\rmatO}{\bimO}
\safemath{\rmatP}{\bimP}
\safemath{\rmatQ}{\bimQ}
\safemath{\rmatR}{\bimR}
\safemath{\rmatS}{\bimS}
\safemath{\rmatT}{\bimT}
\safemath{\rmatU}{\bimU}
\safemath{\rmatV}{\bimV}
\safemath{\rmatW}{\bimW}
\safemath{\rmatX}{\bimX}
\safemath{\rmatY}{\bimY}
\safemath{\rmatZ}{\bimZ}

\safemath{\rmatDelta}{\bimDelta}
\safemath{\rmatLambda}{\bimLambda}
\safemath{\rmatPhi}{\bimPhi}
\safemath{\rmatSigma}{\bimSigma}
\safemath{\rmatOmega}{\bimOmega}
\safemath{\rmatTheta}{\bimTheta}

\usepackage{amssymb}
\usepackage{amsfonts}
\usepackage{mathrsfs}
\usepackage{xspace}
\usepackage{bm}
\usepackage{fancyref}
\usepackage{textcomp}

\usepackage{multirow}
\usepackage{stmaryrd}

\newenvironment{textbmatrix}{	\setlength{\arraycolsep}{2.5pt}%
								\big[\begin{matrix}}{\end{matrix}\big]%
								\raisebox{0.08ex}{\vphantom{M}}}

\def\be{\begin{equation}}
\def\ee{\end{equation}}
\def\een{\nonumber \end{equation}}
\def\mat{\begin{bmatrix}}
\def\emat{\end{bmatrix}}
\def\btm{\begin{textbmatrix}}
\def\etm{\end{textbmatrix}}

\def\ba#1\ea{\begin{align}#1\end{align}}
\def\bas#1\eas{\begin{align*}#1\end{align*}}
\def\bs#1\es{\begin{split}#1\end{split}} 
\def\bg#1\eg{\begin{gather}#1\end{gather}}
\def\bml#1\eml{\begin{multline}#1\end{multline}}
\def\bi#1\ei{\begin{itemize}#1\end{itemize}}

\newcommand{\lefto}{\mathopen{}\left}

\DeclareMathOperator*{\argmin}{arg\;min}		

\newcommand{\vecnorm}[1]{\lefto\lVert#1\right\rVert}		

\safemath{\dirac}{\delta}					
\safemath{\krond}{\dirac}					

\safemath{\upto}{\uparrow}
\safemath{\downto}{\downarrow}
\safemath{\iu}{j}							
\safemath{\ev}{\lambda}						
\safemath{\hilseqspace}{l^{2}}				
\newcommand{\banachfunspace}[1]{\setL^{#1}}	
\safemath{\hilfunspace}{\banachfunspace{2}}	

\safemath{\SNR}{\textsf{SNR}} 				
\safemath{\PAR}{\textsf{PAR}} 				
\safemath{\No}{N_0}							
\safemath{\Es}{E_s}							
\safemath{\Eb}{E_b}							
\safemath{\EbNo}{\frac{\Eb}{\No}}
\safemath{\EsNo}{\frac{\Es}{\No}}

\DeclareMathOperator{\CHop}{\ensuremath{\opH}} 
\safemath{\tvir}{\rndh_{\CHop}}				
\safemath{\tvtf}{\rndl_{\CHop}}				
\safemath{\spf}{\rnds_{\CHop}}				
\safemath{\bff}{H_{\CHop}}					

\safemath{\ircf}{r_{h}}						
\safemath{\tftvcf}{r_{s}}					
\safemath{\tfcf}{r_{l}}						
\safemath{\bfcf}{r_{H}}						

\safemath{\tcorr}{c_h}						
\safemath{\scf}{c_{s}}						
\safemath{\tfcorr}{c_{l}}					
\safemath{\fcorr}{c_{H}}						

\safemath{\mi}{I}							
\safemath{\capacity}{C}						

\safemath{\normal}{\mathcal{N}}			
\safemath{\jpg}{\mathcal{CN}}			
\safemath{\mchain}{\leftrightarrow}		

\safemath{\dB}{\,\mathrm{dB}}
\safemath{\dBm}{\,\mathrm{dBm}}
\safemath{\Hz}{\,\mathrm{Hz}}
\safemath{\kHz}{\,\mathrm{kHz}}
\safemath{\MHz}{\,\mathrm{MHz}}
\safemath{\GHz}{\,\mathrm{GHz}}
\safemath{\s}{\,\mathrm{s}}
\safemath{\ms}{\,\mathrm{ms}}
\safemath{\mus}{\,\mathrm{\text{\textmu}s}}
\safemath{\ns}{\,\mathrm{ns}}
\safemath{\ps}{\,\mathrm{ps}}
\safemath{\meter}{\,\mathrm{m}}
\safemath{\mm}{\,\mathrm{mm}}
\safemath{\cm}{\,\mathrm{cm}}
\safemath{\m}{\,\mathrm{m}}
\safemath{\W}{\,\mathrm{W}}
\safemath{\mW}{\, \mathrm{mW}}
\safemath{\J}{\,\mathrm{J}}
\safemath{\K}{\,\mathrm{K}}
\safemath{\bit}{\,\mathrm{bit}}
\safemath{\nat}{\,\mathrm{nat}}

\safemath{\define}{\triangleq}			

\safemath{\equivalent}{\sim}
\safemath{\distas}{\sim}					
\safemath{\sdiff}{\Delta}				

\safemath{\reals}{\mathbb{R}}
\safemath{\positivereals}{\reals_{+}}
\safemath{\integers}{\mathbb{Z}}
\safemath{\posint}{\integers_{+}}
\safemath{\naturals}{\mathbb{N}}
\safemath{\posnaturals}{\naturals_{+}}
\safemath{\complexset}{\mathbb{C}}
\safemath{\rationals}{\mathbb{Q}}

\newcommand*{\fancyrefapplabelprefix}{app}		
\newcommand*{\fancyrefthmlabelprefix}{thm}		
\newcommand*{\fancyreflemlabelprefix}{lem}		
\newcommand*{\fancyrefcorlabelprefix}{cor}		
\newcommand*{\fancyrefdeflabelprefix}{def}		
\newcommand*{\fancyrefproplabelprefix}{prop}	
\newcommand*{\fancyrefobslabelprefix}{obs}		
\newcommand*{\fancyrefalglabelprefix}{alg}		
\newcommand*{\fancyrefasmlabelprefix}{asm}	    
\newcommand*{\fancyreftbllabelprefix}{tbl}	    

\frefformat{vario}{\fancyrefseclabelprefix}{Section~#1}
\frefformat{vario}{\fancyrefthmlabelprefix}{Theorem~#1}
\frefformat{vario}{\fancyreflemlabelprefix}{Lemma~#1}
\frefformat{vario}{\fancyrefcorlabelprefix}{Corollary~#1}
\frefformat{vario}{\fancyrefdeflabelprefix}{Definition~#1}
\frefformat{vario}{\fancyrefobslabelprefix}{Observation~#1}
\frefformat{vario}{\fancyrefasmlabelprefix}{Assumption~#1}
\frefformat{vario}{\fancyreffiglabelprefix}{Figure~#1}
\frefformat{vario}{\fancyrefapplabelprefix}{Appendix~#1} 
\frefformat{vario}{\fancyrefproplabelprefix}{Proposition~#1}
\frefformat{vario}{\fancyrefalglabelprefix}{Algorithm~#1}
\frefformat{vario}{\fancyrefeqlabelprefix}{(#1)}
\frefformat{vario}{\fancyreftbllabelprefix}{Table~#1}

\safemath{\dictab}{[\,\dicta\,\,\dictb\,]}

\safemath{\ysig}{\bmy}
\safemath{\ysighat}{\hat{\ysig}}
\safemath{\ysigdim}{M}
\safemath{\xsig}{\bmx}
\safemath{\xsigdim}{N}
\safemath{\nx}{n_x}
\safemath{\zsig}{\bmz}
\safemath{\zsigdim}{\ysigdim}
\safemath{\rsig}{\bmr}
\safemath{\Adict}{\bA}
\safemath{\Adicttilde}{\widetilde{\Adict}}
\safemath{\Adictdim}{\outputdim\times\xsigdim}
\safemath{\avec}{\bma}
\safemath{\avectilde}{\tilde{\avec}}
\safemath{\Bdict}{\bB}
\safemath{\Bdicttilde}{\widetilde{\Bdict}}
\safemath{\Cdict}{\bC}
\safemath{\cvec}{\bmc}
\safemath{\Ddict}{\bD}
\safemath{\Ddictdim}{\ysigdim\times\xsigdim}
\safemath{\dvec}{\bmd}
\safemath{\Ddicttilde}{\widetilde{\bD}}
\safemath{\Bonb}{\bB}
\safemath{\bvec}{\bmb}
\safemath{\Bonbdim}{\ysigdim\times\ysigdim}
\safemath{\noise}{\bmn}
\safemath{\noisedim}{\ysigim}
\safemath{\err}{\bme}
\safemath{\errdim}{\ysigdim}
\safemath{\errset}{\setE}
\safemath{\nerr}{n_e}
\safemath{\delop}{\bP_\errset}
\safemath{\delopc}{\bP_{{\errset}^c}}

\safemath{\cplxi}{\imath}
\safemath{\cplxj}{\jmath}

\safemath{\dict}{\matD}
\safemath{\inputdim}{N}		
\safemath{\outputdim}{M}		
\safemath{\sparsity}{S}	
\safemath{\inputdimA}{{N_a}}	
\safemath{\inputdimB}{{N_b}}	
\safemath{\elemA}{{n_a}}	
\safemath{\elemB}{{n_b}}	
\safemath{\resA}{\matR_a}	
\safemath{\resB}{\matR_b}	
\safemath{\subD}{\matS} 
\safemath{\subA}{\matS_a} 
\safemath{\subB}{\matS_b} 
\safemath{\dicta}{\matA} 	
\safemath{\dictb}{\matB} 	
\safemath{\hollowS}{H}
\safemath{\hollowA}{H_a}
\safemath{\hollowB}{H_b}
\safemath{\cross}{Z}
\safemath{\coh}{\mu_d}			
\safemath{\coha}{\mu_a}			
\safemath{\cohb}{\mu_b}			
\safemath{\mubs}{\nu}	
\safemath{\cohm}{\mu_m} 
\safemath{\dictset}{\setD}	
\safemath{\dictsetp}{\dictset(\coh,\coha,\cohb)}	
\safemath{\dictsetgen}{\dictset_\text{gen}}
\safemath{\dictsetgenp}{\dictsetgen(\coh)}
\safemath{\dictsetonb}{\dictset_\text{onb}}
\safemath{\dictsetonbp}{\dictsetonb(\coh)}

\safemath{\leftside}{U}
\safemath{\rightsideA}{R_a}
\safemath{\rightsideB}{R_b}

\safemath{\indexS}{\setI_S} 

\safemath{\na}{n_a}			
\safemath{\nb}{n_b}			
\safemath{\coeffa}{p_i}	
\safemath{\coeffb}{q_j}	
\safemath{\seta}{\setP}		
\safemath{\setb}{\setQ}     
\safemath{\setw}{\setW}	
\safemath{\setz}{\setZ}	
\safemath{\cola}{\veca}		
\safemath{\colb}{\vecb}		
\safemath{\cold}{\vecd}		
\safemath{\inputvec}{\vecx} 	
\safemath{\error}{\vece}	
\safemath{\noiseout}{\vecz} 	
\safemath{\inputvecel}{x}
\safemath{\inputveca}{\vecx_a}
\safemath{\inputvecb}{\vecx_b}
\safemath{\outputvec}{\vecy}	
\safemath{\lambdamin}{\lambda_{\mathrm{min}}}

\safemath{\elltwo}{\ell_2}
\safemath{\ellone}{\ell_1}
\safemath{\ellzero}{\ell_0}
\safemath{\ellinf}{\ell_\infty}
\safemath{\ellinftilde}{\ell_{\widetilde\infty}}
\safemath{\licard}{Z(\coh,\coha,\cohb)}
\safemath{\xsol}{\hat{x}}
\safemath{\xbord}{x_b}		
\safemath{\xstat}{x_s}		
\safemath{\xstatLone}{\tilde{x}_s}
\safemath{\order}{\mathcal{O}} 
\safemath{\scales}{\Theta} 
\safemath{\ones}{\mathbf{1}} 
\safemath{\zeroes}{\mathbf{0}} 
\safemath{\thlone}{\kappa(\coh,\cohb)} 
\safemath{\constoneA}{\delta} 
\safemath{\constoneB}{\epsilon} 
\safemath{\nlarge}{L}				   
\safemath{\sumlarge}{S_\nlarge}
\safemath{\maxlarger}{P_\nlarge}	   
\safemath{\Pzero}{\textrm{P0}}	
\safemath{\Pone}{\textrm{P1}}
\safemath{\vecfir}{\vecw}			 
\safemath{\vecsec}{\vecz}
\safemath{\elvecfir}{w}              
\safemath{\elvecsec}{z}				 
\safemath{\nlargefir}{n}
\safemath{\normout}{\gamma}
\safemath{\auxfun}{h}
\safemath{\supp}{\textrm{supp}}

\safemath{\indexa}{\ell}
\safemath{\indexb}{r}
\safemath{\indexc}{i}
\safemath{\indexd}{j}

\safemath{\project}{P}

\newcommand{\OPP}{{\text{OPP}}}

\newtheorem{alg}{Algorithm}

\IEEEoverridecommandlockouts 
\allowdisplaybreaks 
\usepackage{color}

\graphicspath{{./figs/}}

\begin{document}

\title{Neural-Network Optimized 1-bit Precoding\\for Massive MU-MIMO\\[-0.1cm]}
\author{\IEEEauthorblockN{Alexios Balatsoukas-Stimming\IEEEauthorrefmark{1}\IEEEauthorrefmark{2}, Oscar Casta\~neda\IEEEauthorrefmark{2}, Sven Jacobsson\IEEEauthorrefmark{3}\IEEEauthorrefmark{4}, Giuseppe Durisi\IEEEauthorrefmark{4}, and Christoph Studer\IEEEauthorrefmark{2}} \\[-0.2cm]%
\IEEEauthorblockA{
\small
\IEEEauthorrefmark{1}\textit{Telecommunications Circuits Laboratory, \'Ecole Polytechnique F\'ed\'erale de Lausanne, Switzerland; email: alexios.balatsoukas@epfl.ch}\\
\IEEEauthorrefmark{2}\textit{School of Electrical and Computer Engineering, Cornell University, Ithaca, NY; email: oc66@cornell.edu, studer@cornell.edu}\\
\IEEEauthorrefmark{3}\textit{Ericsson Research, Gothenburg, Sweden; email: sven.jacobsson@ericsson.com}\\
\IEEEauthorrefmark{4}\textit{Chalmers University of Technology, Gothenburg, Sweden; email: durisi@chalmers.se\vspace{-0.3cm}}
\thanks{The work of ABS was supported by the Swiss NSF under projects \#184772 and \#182621.
The work of OC and CS was supported in part by Xilinx Inc. and by the US NSF under grants ECCS-1408006, CCF-1535897,  CCF-1652065, CNS-1717559, and ECCS-1824379. 
}
}
}

\maketitle


\begin{abstract}
Base station (BS) architectures for massive multi-user (MU) multiple-input multiple-output (MIMO) wireless systems are equipped with hundreds of antennas to serve tens of users on the same time-frequency channel. The immense number of BS antennas incurs high system costs, power, and interconnect bandwidth. To circumvent these obstacles, sophisticated MU precoding algorithms that enable the use of 1-bit DACs have been proposed. Many of these precoders feature parameters that are, traditionally, tuned manually to optimize their performance. We propose to use deep-learning tools to automatically tune such 1-bit precoders. Specifically, we optimize the biConvex 1-bit PrecOding (C2PO) algorithm using neural networks. Compared to the original C2PO algorithm, our neural-network optimized (NNO-)C2PO achieves the same error-rate performance at $\bf 2\boldsymbol\times$ lower complexity. Moreover, by training NNO-C2PO for different channel models, we show that 1-bit precoding can be made robust to vastly changing propagation conditions. 
\end{abstract}




\section{Introduction}
\label{sec:introduction}
Massive multiuser (MU) multiple-input multiple-output (MIMO) will be a core technology in fifth-generation (5G) wireless systems~\cite{rusek14a, larsson14a, lu14a}. 
By equipping the base station (BS) with hundreds of antennas, massive MU-MIMO permits communication with tens of user equipments (UEs) in the same time-frequency resource using fine-grained beamforming.
However, such systems with hundreds of BS antennas, in which each antenna is connected to a high-precision radio frequency (RF) chain, would result in prohibitively high system costs, power consumption, and interconnect bandwidth between the baseband processing unit and the remote radios. 
Thus, novel BS architectures, in combination with efficient baseband-processing algorithms, are necessary to reduce the cost, power, and interconnect bandwidth of massive MU-MIMO systems, while preserving high reliability and spectral efficiency.

\subsection{Massive MU-MIMO with Low-Precision BS Architectures}
The cost, power, and interconnect bandwidth of massive MU-MIMO BSs can be greatly reduced by using low-precision digital-to-analog converters (DACs).
Such an architecture tolerates decreased linearity and noise requirements for the RF circuitry, enabling the use of low-cost and power-efficient analog circuits.
However, advanced baseband algorithms are needed to attain high spectral efficiency.

In the downlink (i.e., when the BS transmits to the UEs), linear precoders (e.g., maximal-ratio transmission (MRT) or zero-forcing (ZF) precoding) followed by quantization result in low complexity algorithms that suppress MU interference using low-precision DACs, at the cost of a significant bit error-rate (BER) performance degradation~\cite{mezghani09c,saxena16b,li17a}.
In contrast, sophisticated nonlinear precoders~\cite{jacobsson17d, jacobsson16b,jedda16a,tirkkonen17a,swindlehurst17a,castaneda17b,castaneda2017cxpo,shao18a,nedelcu17a,jedda18b} achieve superior BER performance (especially in the extreme case where a pair of $1$-bit DACs per RF chain is used), at the cost of a higher complexity.
Existing nonlinear precoding algorithms require manual parameter tuning to optimize the BER performance~\cite{castaneda2017cxpo}. Parameter tuning is time-consuming and may yield suboptimal results in realistic systems with vastly changing propagation conditions.
Furthermore, most existing precoders have only been evaluated in idealistic Rayleigh-fading conditions, and it remains unclear whether they also provide a satisfactory performance in realistic channels. 
%
%

\subsection{Contributions}
In this paper, we propose to use neural-network optimization tools to tune the algorithm parameters of the nonlinear biConvex 1-bit PrecOding (C2PO) algorithm proposed in~\cite{castaneda2017cxpo}.
Specifically, we unfold the C2PO iterations and use backpropagation for parameter tuning.
We show that the resulting neural-network optimized C2PO (NNO-C2PO) algorithm delivers significantly improved performance when compared to C2PO for the same number of iterations, for a range of channel models.
%
%
More interestingly, we show that NNO-C2PO can use the same set of learned per-iteration parameter values on vastly different channel models at a negligible performance loss compared to the case where the algorithm is separately trained for each channel model.
In contrast, the original C2PO algorithm is unable to deal with different channel models, even when its single set of parameters is trained in the same way as NNO-C2PO.
Overall, our results illustrate that, by using machine learning techniques, existing iterative nonlinear precoders can significantly improve their BER performance, while also becoming robust to changing channel conditions.

\subsection{Relevant Prior Art}
Several recent works have explored the application of machine-learning techniques to different problems that arise in the context of massive MU-MIMO.
The majority of results have focused on data detection \cite{samuel2017a, klautau2018a, corlay2018a, khobahi18a, he2018a, takabe18a, kim19a}, while some have tackled channel estimation \cite{borgerding2017a, klautau2018a}, and only~\cite{xia19a} has considered the case of (non-quantized) downlink precoding. 
However, most of these results~\cite{samuel2017a,klautau2018a,corlay2018a,khobahi18a,xia19a} consider deep learning models that are highly parametrizable (e.g., containing multiple dense layers), and thus, both difficult to train and too complex to implement in practice. 
%
%
The methods proposed in~\cite{he2018a,takabe18a} are the most closely related to our work, since they also unfold iterative algorithms and learn a small set of algorithm parameters from training data. We note that some of the parameters that are trained in \cite{he2018a,takabe18a} are introduced arbitrarily into the algorithm, while the C2PO algorithm that we use already contains tunable parameters by design.
To our knowledge, the unfolding technique has not been examined in the context of 1-bit precoding in massive MU-MIMO systems before. Moreover, the robustness of the learned communications systems (particularly, 1-bit massive MU-MIMO systems) to changing channel conditions is a practically-relevant issue that has not been addressed to date.

\section{System Model and $1$-bit Precoding}
\label{sec:model}

\setlength{\textfloatsep}{10pt}
\begin{figure}[t]
\centering
\includegraphics[width=0.95\columnwidth]{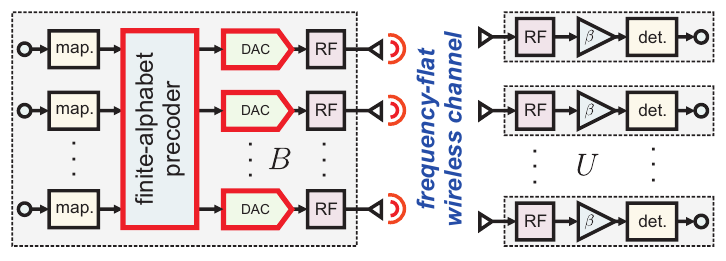}
\vspace{-0.3cm}
 \caption{Simplified view of a massive MU-MIMO downlink system with low-resolution DACs: A BS with $B$ antennas uses a finite-alphabet precoder to transmit data to $U$ single-antenna UEs through a wireless channel.} 
\label{fig:system_overview}
\end{figure}

\subsection{System Model}
We focus on the downlink of a single-cell, narrowband massive MU-MIMO system as shown in \fref{fig:system_overview}, where a BS equipped with $B$ antennas serves $U \ll B$ UEs with one antenna each.
This system is modeled as $\vecy = \matH \vecx + \vecn$, where $\vecy = [y_1,\dots,y_U]^T\in\complexset^U$ comprises the signals received by each UE, $\matH \in \opC^{U \times B}$ is the downlink MIMO channel matrix (which is assumed to be perfectly known at the BS), $\bmn\in\complexset^U$ models i.i.d.\ circularly-symmetric complex Gaussian noise with variance~$N_0$ per complex~entry, and $\vecx\in\setX^B$ is the \emph{precoded vector} with entries chosen from the discrete set $\setX$ determined by the DAC outputs.  
We assume that the precoded vector satisfies an instantaneous power constraint $\|\vecx\|_2^2\le P$.

\subsection{$1$-bit Precoding}
%
For the case of $1$-bit precoding, where both the in-phase and quadrature paths are fed by 1-bit DACs, we have $\setX = \{+\upsilon+j\upsilon,+\upsilon-j\upsilon,-\upsilon+j\upsilon,-\upsilon-j\upsilon\}$, where $\upsilon\in\reals_+$ is set to satisfy the power constraint. 
The objective of the precoder at the BS is to take a symbol vector ${\vecs=[s_1,\dots,s_U]^T\in\setO^U}$ and to map it to a precoded vector $\vecx\in\setX^B$ that is transmitted through the BS antennas. The vector $\vecx$ is chosen so that, after transmission through the wireless channel, the received signal $y_u$ at the $u$th UE is proportional to the transmitted symbol $s_u$.
We assume that each UE is able to rescale the received signal $y_u$ by a common scalar $\beta\in\complexset$ to estimate the symbol $s_u$, i.e., the UEs compute $\hat{s}_u=\beta y_u$.
For mean-squared error (MSE)-optimal precoders, the optimal precoding problem (OPP) can then be formulated as~\cite{castaneda2017cxpo}:
\begin{align*}
\text{(\OPP)} \quad 
\{\hat\vecx,\hat\beta\} = \underset{\bmx \in \setX^{B}\!,\, \beta \in \complexset}{\text{arg\,min}} \,\, \vecnorm{\bms - \beta \matH\bmx}^2_2 + |\beta|^2 U \No.
\end{align*}
As shown in~\cite{jacobsson16b}, the users can accurately estimate $\hat{\beta}$.
For 1-bit massive MU-MIMO systems, several algorithms have been proposed to approximately solve (OPP) using convex \cite{jacobsson17d,jacobsson16b} and nonconvex \cite{castaneda2017cxpo} relaxations.
\subsection{C2PO: Biconvex 1-bit Precoding}
In this work, we will follow the C2PO algorith put forward in \cite{castaneda2017cxpo}, as it leads to a simple, yet efficient, algorithm that has been implemented in hardware.
To arrive at low-complexity approximate solutions to (OPP), C2PO assumes that the system operates in the high-SNR regime, so that $N_0\to0$.
%
%
Then, with an additional approximation~\cite[Eq.~(2)]{castaneda2017cxpo}, (OPP) is re-written:
\begin{align*}
\text{(\OPP$^*$)} \qquad 
\{\hat{\bmx}, \hat\alpha\} = \argmin_{\bmx \in \setX^{B},\, \alpha \in \complexset}  \vecnorm{\alpha{\vecs} -  {\matH}{\vecx}}^2_2.
\end{align*}
The optimal value for $\alpha$, can be computed in closed form as $\hat{\alpha}=  {{\bms}^H{\bH}{\bmx}}/{\|{\bms}\|^2_2}$. 
%
Substituting this result into (\OPP$^*$) leads to the following equivalent form:
\begin{align*}
\text{(\OPP$^{**}$)} \qquad 
\hat{\bmx} = \argmin_{\bmx \in \setX^{B}} \textstyle \frac{1}{2}\vecnorm{\bA\vecx}^2_2,
\end{align*}
where $\bA =  \left(\bI_U-{\bms\bms^H}/{\|\bms\|^2_2}\right)\bH$. The finite-alphabet constraint $\vecx\in\setX^B$ is then relaxed to the convex hull $\setB$ of the points in $\setX$.
%
%
To avoid the trivial all-zeros solution, a concave regularizer $-\frac{\delta}{2}\|\vecx\|^2_2$ with $\delta>0$ is added to the objective:
%
%
\begin{align}
\hat{\bmx} = \argmin_{\bmx \in \setB^{B}} \,\, \textstyle \frac{1}{2} \vecnorm{\bA\vecx}^2_2    -  \frac{\delta}{2} \|\bmx\|_2^2. \label{eq:directform}
\end{align}

C2PO uses forward-backward splitting (FBS)~\cite{GSB14,BT09} to solve \fref{eq:directform}, an efficient numerical method for convex optimization problems. 
As the objective in \fref{eq:directform} is nonconvex, FBS is not guaranteed to converge to an optimal solution.
However, C2PO has been shown to yield excellent performance in practice \cite{castaneda2017cxpo}.
%
%
After applying FBS to \fref{eq:directform}, the C2PO algorithm is obtained (see \cite{castaneda2017cxpo} for a detailed explanation). To present C2PO, we need to introduce some auxiliary notation. Let  $\rho=\frac{1}{1-\tau\delta}$, $\xi = \sqrt{\frac{P}{2B}}$, and the so called proximal-operator:
\begin{equation}
\small
\mathrm{prox}_g (\bmz;\rho,\xi)\!=\!\mathrm{clip}\!\left(\!\rho\Re\{\bmz\}, \xi\right) + j~\text{clip}\!\left(\!\rho\Im\{\bmz\}, \xi \right),
\label{eq:proxg}
\end{equation}
where the clipping function $\mathrm{clip}(\bmz,\gamma)$ applies the operation $\min(\max(z_i,-\gamma),\gamma)$ to each element of the vector~$\bmz$. 
%
%
\begin{figure*}
	\centering
	\includegraphics[width=0.975\textwidth]{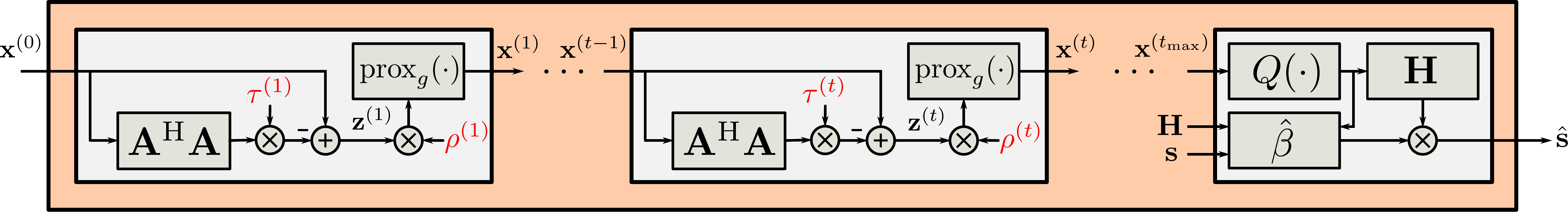}
	\caption{Computation graph of the iteration-unfolded version of NNO-C2PO. All trainable parameters are highlighted in red color.}\label{fig:nnofap}
\end{figure*}
%
%
%
%
\begin{oframed}
\vspace{-0.3cm}
\begin{alg}[C2PO] \label{alg:C2PO} 
Initialize \mbox{$\bmx^{(0)}=\bH^H\bms$}. Fix $\tau^{(t)}$ and $\rho^{(t)}$. For every iteration $t=1,2,\ldots,t_\text{max}$ compute: 
\begin{align}
\bmz^{(t)} &= \bmx^{(t-1)} - \tau^{(t)} \bA^H\bA \bmx^{(t-1)}  \label{eq:faststep1} \\
\bmx^{(t)} & = \mathrm{prox}_g ( {\bmz^{(t)}};\rho^{(t)},\xi), \label{eq:faststep2} 
\end{align}
Finally, quantize the output $\bmx^{(t_\text{max})}$ to the set $\setX^B$.
\end{alg}
\vspace{-0.2cm}
\end{oframed}
We note that in the original reference~\cite{castaneda2017cxpo}, constant values were used for $\tau=\tau^{(t)}$ and $\rho=\rho^{(t)}$ at every iteration because manually tuning $2t_{\max}$ distinct parameters seemed impractical.

\section{Neural-Network Optimized C2PO}
\label{sec:nnofap}

Analogous to~\cite{he2018a,takabe18a}, the idea behind our approach is to unfold the iterations of C2PO and to learn distinct parameters $\tau ^{(t)}$ and $\rho ^{(t)}$ for each iteration $t=1,\ldots,t_{\max}$. These parameters are learned offline using deep-learning methods, and then used in the C2PO algorithm as described in \fref{alg:C2PO}. We refer to the overall procedure as neural-network optimized C2PO (NNO-C2PO).

\subsection{Unfolding the Algorithm}
As shown in \fref{fig:nnofap}, the unfolded algorithm forms a computation graph, which can be described using standard deep learning frameworks (e.g., Keras or PyTorch). If all operations in the computation graph have well-defined gradients, then the gradients of the cost function with respect to $\tau ^{(t)}$ and $\rho ^{(t)}$ can be calculated efficiently (and easily, due to automatic differentiation) using backpropagation. These gradients can then be used to learn the values of $\tau^{(t)}$ and $\rho^{(t)}$. 

The high-level architecture of the unfolded NNO-C2PO is shown in \fref{fig:nnofap}. 
The algorithm takes $\mathbf{H}$, $\bms$, $\bmx^{(0)}$, and $\mathbf{A}^H\mathbf{A}$ as inputs and produces $\hat{\bms}$ as an output. We note that $\bmx^{(0)}$ and $\mathbf{A}$ are functions of $\bms$ and $\mathbf{H}$, but they are pre-computed and given as additional inputs for convenience. The input to the first sub-block is $\bmx^{(0)} = \mathbf{H}^{H}\bms$. Each successive sub-block takes $\bmx^{(t-1)}$ as  input and produces $\bmx^{(t)}$ by means of~\eqref{eq:faststep1} and~\eqref{eq:faststep2}. The final sub-block quantizes $\bmx^{(t_{\max})}$ to the finite alphabet $\setX^B$ as: 
\begin{align}
	\bmx^{(t_{\max})}_Q	& = Q\left(\bmx^{(t_{\max})};\xi\right),
\end{align}
where $Q(\cdot)$ is a 1-bit quantization function that is applied element-wise to $\bmx^{(t_{\max})}$ and is defined as:
\begin{align}
	Q(x;\xi) & = \xi\text{sign}\left(\Re\{x\}\right) + j \cdot \xi\text{sign}\left(\Im\{x\}\right).
\end{align}
Moreover, the final block also computes the scalar value $\hat{\beta} = {\|{\bms}\|^2_2}/{{\bms}^H{\bH}{\bmx^{(t_{\max})}_Q}}$, and emulates the (noiseless, since by assumption $N_0 \to 0$) transmission over the channel as:
\begin{align}
	\hat{\bms}	& = \hat{\beta} \cdot \mathbf{H}\bmx^{(t_{\max})}_Q.
\end{align}


\subsection{Implementation Details}

All operations involved in~\eqref{eq:faststep1} are linear and thus have well-defined gradients. However, care needs to be taken for the operations  in~\eqref{eq:faststep2} as well as in the last stage in \fref{fig:nnofap}. More specifically, in~\eqref{eq:faststep2}, the gradient of the function $\mathrm{clip}(\bmz,\gamma)$ that is applied element-wise to $\bmz^{(t)}$ has a discontinuity for $z = \gamma$ and $z = -\gamma$, but since the gradients are evaluated numerically, this is not an issue in practice. Moreover, the gradient of $\mathrm{clip}(z,\gamma)$ with respect to $z$ is equal to $0$ for $z < -\gamma$ and $z > \gamma$. However, as $\mathrm{clip}(z,\gamma)$ is applied to $B$ elements of $\bmz^{(t)}$ concurrently and over several training samples, the probability that the (averaged) gradient of $\mathrm{clip}(\bmz^{(t)},\gamma)$ with respect to $\rho^{(t)}$ is equal to $0$ is very low. 

The quantization function $Q(\cdot)$ used in the last stage in \fref{fig:nnofap} is not differentiable. There are several ways to include quantization functions into deep-learning frameworks, such as using a soft sign function~\cite{samuel2017a} or a $\tanh(\cdot)$ function~\cite{takabe18a}. Here, we follow the approach used by binarized neural networks (BNNs)~\cite{hubara2016bnns}, which applies $Q(\cdot)$ only during the forward propagation, and replaces $Q(x;\xi)$ with a clipping function $\text{clip}(x;\xi)$ (as a straight-through estimator) during backpropagation. All remaining operations in the last stage of \fref{fig:nnofap} have well-defined gradients and can be differentiated automatically.

We note that the updates in~\eqref{eq:faststep1} and~\eqref{eq:faststep2} are performed in the complex domain.  Most deep learning tools are unable  to deal with  complex numbers. However, all complex operations in~\eqref{eq:faststep1} and~\eqref{eq:faststep2}  can be easily recasted as real-valued operations. 
For example, the operation $\bmx = \mathbf{H}\bms$ can be equivalently performed in the real domain by setting:
\begin{align}
	\mathbf{H}^{\mathbb{R}}	& = \begin{bmatrix} \Re\{\mathbf{H}\}	& - \Im\{\mathbf{H}\} \\ \Im\{\mathbf{H}\} & \Re\{\mathbf{H}\}\end{bmatrix} \quad \text{and} \quad \bms^{\mathbb{R}} = \begin{bmatrix} \Re\{\bms\} \\ \Im\{\bms\} \end{bmatrix},
\end{align}
computing $\bmx^{\mathbb{R}} = \mathbf{H}^{\mathbb{R}}\bms^{\mathbb{R}}$ and, finally, setting:
\begin{align}
	x_b = x^{\mathbb{R}}_{b} + j~x^{\mathbb{R}}_{b+B}, \quad b = 1,\hdots,B.
\end{align}

\subsection{Training}

When unfolding the iterations of an iterative algorithm, the corresponding computation graph can become very deep. As such, the well-known problem of vanishing gradients may make training difficult. 
This problem was identified in~\cite{takabe18a} in the context of massive MU-MIMO and an incremental training procedure was used to enable effective training. 
However, since we unfold only a small number of iterations in our experiments and we only learn a small set of parameters, we did not encounter this problem.
Hence, performing a one-shot training of the entire network suffices. 
The training set consists of (the real-valued equivalents of) $K$ training samples $\left\{\bms_{(k)},\mathbf{H}_{(k)}, \bmx^{(0)}_{(k)},\mathbf{A}^H_{(k)}\mathbf{A}_{(k)}\right\}$, where $k = 1,\hdots,K$. The desired output is $\bms_{(k)}$ and the cost function to be minimized is the MSE between $\bms_{(k)}$ and $\hat{\bms}_{(k)}$:
\vspace{-0.1cm}
\begin{align}
	C = \frac{1}{K}\sum_{k=1}^K\left(\vecnorm{\bms_{(k)}-\hat{\bms}_{(k)}}_2^2\right). \label{eq:cost}
\end{align}
Training the network essentially means choosing values for $\tau ^{(t)}$ and $\rho ^{(t)}$ that minimize~\eqref{eq:cost}. The parameters are initialized to $\tau ^{(t)} = 2^{-8}$ and  $\rho ^{(t)} = 1.25$, which have been shown to work well in practice~\cite{castaneda2017cxpo}.

\section{Results}
\label{sec:results}

We now  compare the performance of the proposed NNO-C2PO algorithm with the C2PO algorithm. For C2PO, we learn its parameters using the  same procedure as for NNO-C2PO, but we constrain the weights so that $\tau^{(t)} = \tau$ and $\rho^{(t)} = \rho$ for all iterations $t = 1,\hdots,t_{\max}$. We provide BER results for Rayleigh-fading channels, as well as for more realistic line-of-sight (LoS) and non-line-of-sight (NLoS) channels generated with the QuaDRiGa channel model~\cite{jaeckel2014quadriga} using transmission at 2GHz with a 6.8kHz bandwidth in the ``Berlin UMa'' scenario. For all experiments, we consider a massive MU-MIMO system with $B = 128$ BS antennas, $U = 8$ UEs, and $16$-QAM.
We generate $K=500$ training samples for each channel model and we perform full batch training over $100t_{\max}$ epochs. The BER performance evaluation is carried out using $K_{\text{eval}}=1000$ evaluation samples that are distinct from the training samples. The computation graph of \fref{fig:nnofap} is described using  Keras~\cite{chollet2015keras} with a TensorFlow backend \cite{tensorflow2015-whitepaper}. We use the Adam optimizer~\cite{kingma2015adam} with a learning rate of $\lambda = 10^{-4}$.

\subsection{Rayleigh-Fading Channel}
\begin{figure}[t]
	\centering
	\includegraphics[width=0.9\columnwidth]{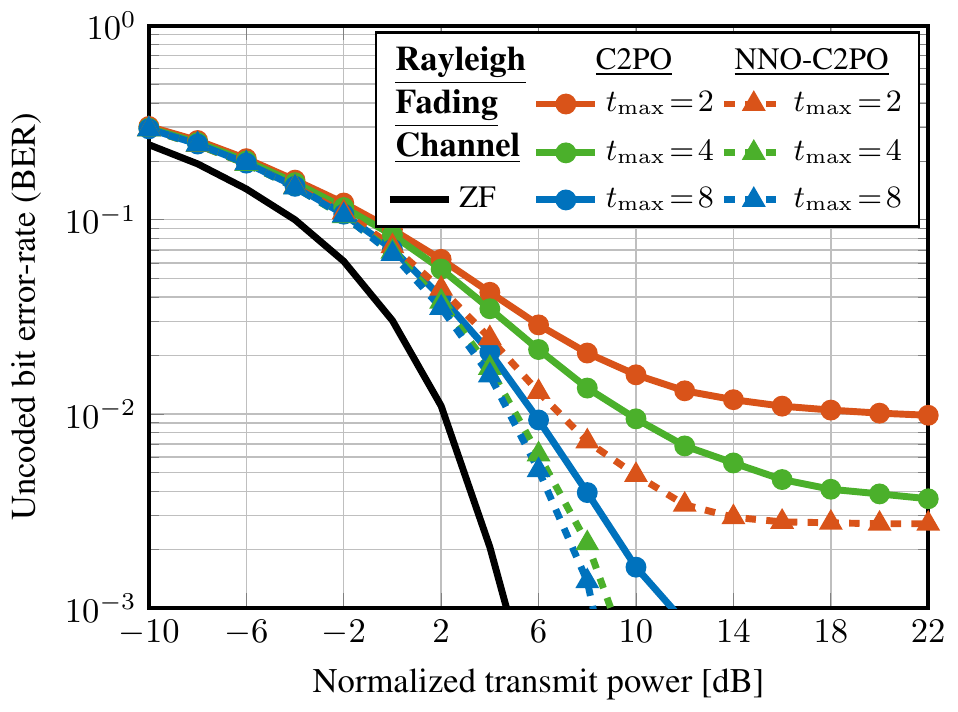}
	\caption{Uncoded bit error-rate (BER) performance of C2PO and NNO-C2PO, using different number of iterations $t_{\max}$, for a massive MU-MIMO system with $B\!=\!128$ BS antennas and $U\!=\!8$ UEs, operating with $16$-QAM in a Rayleigh-fading channel. Zero-forcing (ZF) precoding is included as reference.}
	\label{fig:BER}
	\vspace{-0.2cm}
\end{figure}
The BER performance of NNO-C2PO is compared to that of C2PO under Rayleigh-fading and for different values of $t_{\max}$ in \fref{fig:BER}, where we also show the infinite-precision ZF solution for reference. We observe that NNO-C2PO with $t_{\max} = 2$ and $t_{\max} = 4$ slightly outperforms C2PO with $t_{\max} = 4$ and $t_{\max} = 8$, respectively. This means that NNO-C2PO requires approximately $50\%$ fewer iterations than C2PO to achieve similar BER performance. Moreover, even when considering the case of $t_{\max} = 8$ where NNO-C2PO delivers diminishing returns, there is a difference of roughly $3$~dB between the two algorithms at a BER of $0.1\%$. Note that this improvement comes at essentially no cost, since the C2PO hardware architecture of~\cite{castaneda2017cxpo} can be modified to support NNO-C2PO with very little hardware overhead to operate with the per-iteration values of $\tau^{(t)}$ and $\rho^{(t)}$.

\begin{figure}[t]
	\centering
	\includegraphics[width=0.87\columnwidth]{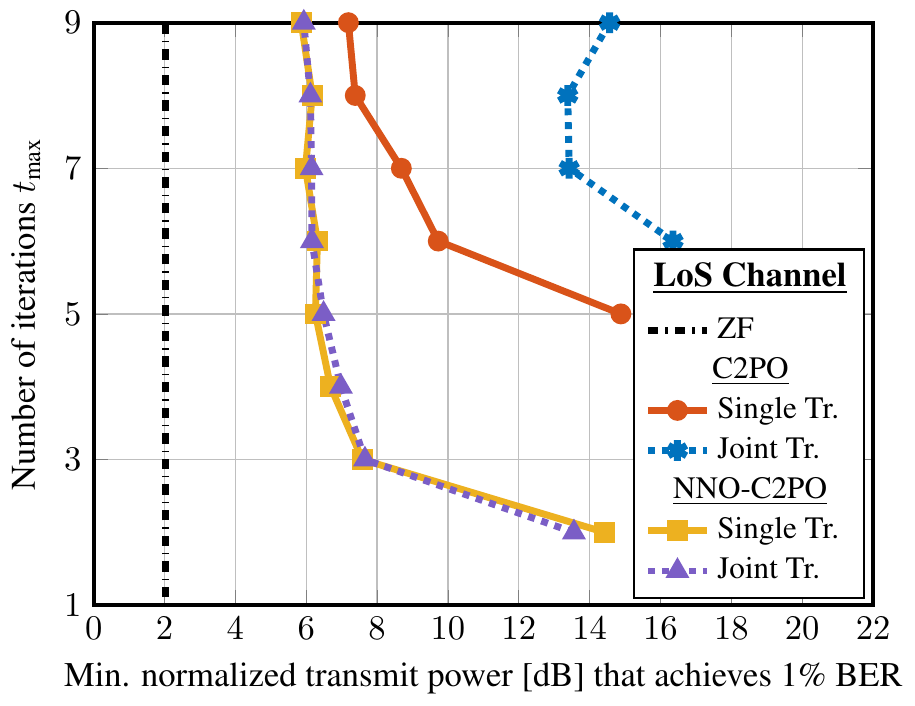}
	\caption{Performance-iteration trade-offs for C2PO and NNO-C2PO for a massive MU-MIMO system with $B\!=\!128$ BS antennas and $U\!=\!8$ UEs, operating with $16$-QAM in a realistic LoS channel. The vertical line shows the normalized transmit power at which ZF precoding achieves $1\%$ BER.}
	\label{fig:LOS}
	\vspace{-0.2cm}
\end{figure}

\begin{figure}[t]
	\centering
	\includegraphics[width=0.87\columnwidth]{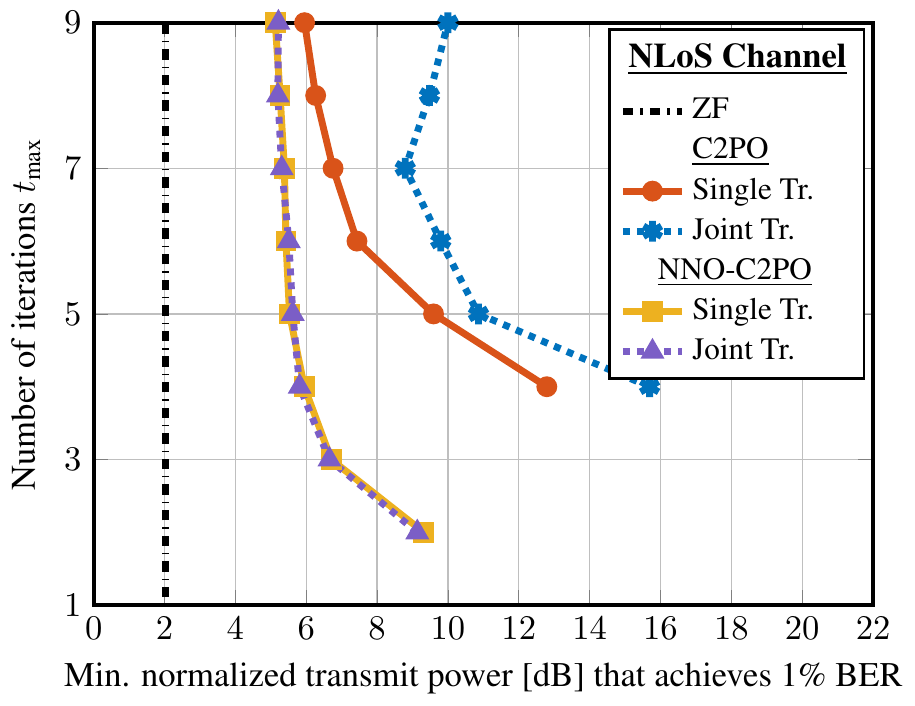}
	\caption{Performance-iteration trade-offs for C2PO and NNO-C2PO for a massive MU-MIMO system with $B\!=\!128$ BS antennas and $U\!=\!8$ UEs, operating with $16$-QAM in a realistic NLoS channel. The vertical line shows the normalized transmit power at which ZF precoding achieves $1\%$ BER.}
	\label{fig:NLOS}
	\vspace{-0.2cm}
\end{figure}

\subsection{QuaDRiGa LoS and NLoS Channels}
For the QuaDRiGa LoS and NLoS channel models, we present performance results using a different metric. Specifically, we fix the normalized transmit power, and then we determine the minimum number of C2PO and NNO-C2PO iterations that are required to achieve a target BER of $1\%$.
Moreover, in order to examine the robustness of C2PO and NNO-C2PO to different channel conditions, we provide results for two scenarios. In the \emph{single training} scenario, C2PO and NNO-C2PO are trained on a dataset that contains training data from only one channel model, and they are tested on data from the same channel model they were trained for. In the \emph{joint training} scenario, C2PO and NNO-C2PO are trained on a dataset that contains training data from both channel models, but they are tested on each channel model separately. 

The results of the above experiments are shown in Figures~\ref{fig:LOS} and~\ref{fig:NLOS}, for the LoS and NLoS channels, respectively. We observe that, in the single training scenario, NNO-C2PO provides significant gains with respect to C2PO for both channel models. For example, on the LoS and NLoS channels, NNO-C2PO with $t_{\max} = 5$ achieves the target BER at $9$~dB and $4$~dB lower transmit power than C2PO, respectively. More interestingly, we observe that in the joint training scenario, the performance of C2PO on both the LoS and the NLoS channels becomes significantly worse, while the performance of NNO-C2PO remains nearly the same. The fact that NNO-C2PO is more parameterizable seems to make it sufficiently robust so that it can learn to operate on vastly different channel models with the same set of parameters $\{\tau^{(t)}, \rho^{(t)}\}$.

\section{Conclusions}
\label{sec:conclusions}

We have shown that introducing per-iteration parameters to the C2PO precoding algorithm for 1-bit massive MU-MIMO systems and tuning them using deep-learning tools, enables significantly improved error-rate performance compared to a similarly-tuned C2PO variant that uses the same parameters for all iterations.
Such improved performance is achieved without increasing the algorithmic complexity.
Furthermore, by learning the algorithm parameters for different channel models, we have observed that our NNO-C2PO algorithm is robust to vastly changing propagation conditions, unlike its traditional C2PO counterpart.
Our results indicate that tedious manual parameter tuning of nonlinear 1-bit precoding algorithms can be avoided in practice, while significantly improving their performance at the same time.


\bibliographystyle{IEEEtran} 

\bibliography{IEEEabrv,confs-jrnls,publishers,studer,svenbib,tom}

\begin{thebibliography}{10}
\providecommand{\url}[1]{#1}
\csname url@samestyle\endcsname
\providecommand{\newblock}{\relax}
\providecommand{\bibinfo}[2]{#2}
\providecommand{\BIBentrySTDinterwordspacing}{\spaceskip=0pt\relax}
\providecommand{\BIBentryALTinterwordstretchfactor}{4}
\providecommand{\BIBentryALTinterwordspacing}{\spaceskip=\fontdimen2\font plus
\BIBentryALTinterwordstretchfactor\fontdimen3\font minus
  \fontdimen4\font\relax}
\providecommand{\BIBforeignlanguage}[2]{{%
\expandafter\ifx\csname l@#1\endcsname\relax
\typeout{** WARNING: IEEEtran.bst: No hyphenation pattern has been}%
\typeout{** loaded for the language `#1'. Using the pattern for}%
\typeout{** the default language instead.}%
\else
\language=\csname l@#1\endcsname
\fi
#2}}
\providecommand{\BIBdecl}{\relax}
\BIBdecl

\bibitem{rusek14a}
F.~Rusek, D.~Persson, B.~Kiong, E.~G. Larsson, T.~L. Marzetta, O.~Edfors, and
  F.~Tufvesson, ``Scaling up {MIMO}: Opportunities and challenges with very
  large large arrays,'' \emph{{IEEE} Signal Process. Mag.}, vol.~30, no.~1, pp.
  40--60, Jan. 2013.

\bibitem{larsson14a}
E.~G. Larsson, F.~Tufvesson, O.~Edfors, and T.~L. Marzetta, ``Massive {MIMO}
  for next generation wireless systems,'' \emph{{IEEE} Commun. Mag.}, vol.~52,
  no.~2, pp. 186--195, Feb. 2014.

\bibitem{lu14a}
L.~Lu, G.~Ye~Li, A.~L. Swindlehurst, A.~Ashikhmin, and R.~Zhang, ``An overview
  of massive {MIMO}: Benefits and challenges,'' \emph{{IEEE} J. Sel. Topics
  Signal Process.}, vol.~8, no.~5, pp. 742--758, Oct. 2014.

\bibitem{mezghani09c}
A.~Mezghani, R.~Ghiat, and J.~A. Nossek, ``Transmit processing with low
  resolution {D/A}-converters,'' in \emph{Proc. IEEE Int. Conf. Electron.,
  Circuits, Syst. (ICECS)}, Dec. 2009, pp. 683--686.

\bibitem{saxena16b}
A.~K. Saxena, I.~Fijalkow, and A.~L. Swindlehurst, ``Analysis of one-bit
  quantized precoding for the multiuser massive {MIMO} downlink,'' \emph{{IEEE}
  Trans. Signal Process.}, vol.~65, no.~17, pp. 4624--4634, Sep. 2017.

\bibitem{li17a}
Y.~Li, C.~Tao, A.~L. Swindlehurst, A.~Mezghani, and L.~Liu, ``Downlink
  achievable rate analysis in massive {MIMO} systems with one-bit {DACs},''
  \emph{{IEEE} Commun. Lett.}, vol.~21, no.~7, pp. 1669--1672, Jul. 2017.

\bibitem{jacobsson17d}
S.~Jacobsson, G.~Durisi, M.~Coldrey, T.~Goldstein, and C.~Studer, ``Quantized
  precoding for massive {MU-MIMO},'' \emph{{IEEE} Trans. Commun.}, vol.~65,
  no.~11, pp. 4670--4684, Nov. 2017.

\bibitem{jacobsson16b}
------, ``Nonlinear 1-bit precoding for massive {MU-MIMO} with higher-order
  modulation,'' in \emph{Proc. Asilomar Conf. Signals, Syst., Comput.}, Nov.
  2016, pp. 763--767.

\bibitem{jedda16a}
H.~Jedda, J.~A. Nossek, and A.~Mezghani, ``Minimum {BER} precoding in 1-bit
  massive {MIMO} systems,'' in \emph{{IEEE} Sensor Array and Multichannel Sig.
  Proc. Workshop}, Jul. 2016.

\bibitem{tirkkonen17a}
O.~Tirkkonen and C.~Studer, ``Subset-codebook precoding for 1-bit massive
  multiuser {MIMO},'' in \emph{Conf. Inf. Sci. Syst.}, Mar. 2017.

\bibitem{swindlehurst17a}
A.~L. Swindlehurst, A.~K. Saxena, A.~Mezghani, and I.~Fijalkow, ``Minimum
  probability-of-error perturbation precoding for the one-bit massive {MIMO}
  downlink,'' in \emph{Proc. IEEE Intl. Conf. Acoust., Speech, Signal
  Process.}, Mar. 2017, pp. 6483--6487.

\bibitem{castaneda17b}
O.~Casta\~{n}eda, C.~Studer, and T.~Goldstein, ``{POKEMON}: A non-linear
  beamforming algorithm for 1-bit massive {MIMO},'' in \emph{Proc. IEEE Intl.
  Conf. Acoust., Speech, Signal Process.}, Mar. 2017, pp. 3464--3468.

\bibitem{castaneda2017cxpo}
O.~Casta\~{n}eda, S.~Jacobsson, G.~Durisi, M.~Coldrey, T.~Goldstein, and
  C.~Studer, ``1-bit massive {MU}-{MIMO} precoding in {VLSI},'' \emph{IEEE J.
  Emerging Sel. Topics Circuits Syst.}, vol.~7, no.~4, pp. 508--522, Dec. 2017.

\bibitem{shao18a}
M.~Shao, Q.~Li, and W.-K. Ma, ``One-bit massive {MIMO} precoding via minimum
  symbol-error probability design,'' in \emph{Proc. IEEE Intl. Conf. Acoust.,
  Speech, Signal Process.}, Mar. 2018, pp. 3579--3583.

\bibitem{nedelcu17a}
A.~Nedelcu, F.~Steiner, M.~Staudacher, G.~Kramer, W.~Zirwas, R.~Sisava~Ganesan,
  P.~Baracca, and S.~Wesemann, ``Quantized precoding for multi-antenna downlink
  channels with {MAGIQ},'' in \emph{Int. ITG Workshop Smart Antennas}, Mar.
  2017.

\bibitem{jedda18b}
H.~Jedda, A.~Mezghani, A.~L. Swindlehurst, and J.~A. Nossek, ``Quantized
  constant envelope precoding with {PSK} and {QAM} signaling,'' \emph{{IEEE}
  Trans. Wireless Commun.}, vol.~17, no.~12, pp. 8022--8034, Oct. 2018.

\bibitem{samuel2017a}
N.~Samuel, T.~Diskin, and A.~Wiesel, ``Deep {MIMO} detection,'' in \emph{Proc.
  IEEE Intl. Workshop Signal Process. Adv. Wireless Commun.}, Jul. 2017.

\bibitem{klautau2018a}
A.~Klautau, N.~Gonz\'alez-Prelcic, A.~Mezghani, and R.~W. Heath~Jr.,
  ``Detection and channel equalization with deep learning for low resolution
  {MIMO} systems,'' in \emph{Proc. Asilomar Conf. Signals, Syst., Comput.},
  Oct. 2018, pp. 1836--1840.

\bibitem{corlay2018a}
V.~Corlay, J.~J. Boutros, P.~Ciblat, and L.~Brunel, ``Multilevel {MIMO}
  detection with deep learning,'' in \emph{Proc. Asilomar Conf. Signals, Syst.,
  Comput.}, Oct. 2018, pp. 1805--1809.

\bibitem{khobahi18a}
\BIBentryALTinterwordspacing
S.~Khobahi, N.~Naimipour, M.~Soltanalian, and Y.~C. Eldar, ``Deep signal
  recovery with one-bit quantization,'' Nov. 2018. [Online]. Available:
  \url{https://arxiv.org/abs/1812.00797}
\BIBentrySTDinterwordspacing

\bibitem{he2018a}
H.~He, C.-K. Wen, S.~Jin, and G.~Y. Li, ``A model-driven deep learning network
  for {MIMO} detection,'' in \emph{Proc. IEEE Global Conf. Signal Inf.
  Process.}, Nov. 2018, pp. 584--588.

\bibitem{takabe18a}
\BIBentryALTinterwordspacing
S.~Takabe, M.~Imanishi, T.~Wadayama, and K.~Hayashi, ``Trainable projected
  gradient detector for massive overloaded {MIMO} channels: Data-driven tuning
  approach,'' Dec. 2018. [Online]. Available:
  \url{https://arxiv.org/abs/1812.10044}
\BIBentrySTDinterwordspacing

\bibitem{kim19a}
\BIBentryALTinterwordspacing
S.~Kim and S.-N. Hong, ``Semi-supervised learning detector for {MU-MIMO}
  systems with one-bit {ADCs},'' Feb. 2019. [Online]. Available:
  \url{https://arxiv.org/abs/1902.00866}
\BIBentrySTDinterwordspacing

\bibitem{borgerding2017a}
M.~Borgerding, P.~Schniter, and S.~Rangan, ``{AMP}-inspired deep networks for
  sparse linear inverse problems,'' \emph{IEEE Trans. Signal Process.},
  vol.~65, no.~16, pp. 4293--4308, Aug. 2017.

\bibitem{xia19a}
\BIBentryALTinterwordspacing
W.~Xia, G.~Zheng, Y.~Zhu, J.~Zhang, J.~Wang, and A.~Petropulu, ``A deep
  learning framework for optimization of {MISO} downlink beamforming,'' Jan.
  2019. [Online]. Available: \url{https://arxiv.org/abs/1901.00354}
\BIBentrySTDinterwordspacing

\bibitem{GSB14}
\BIBentryALTinterwordspacing
T.~Goldstein, C.~Studer, and R.~G. Baraniuk, ``A field guide to
  forward-backward splitting with a {FASTA} implementation,'' Nov. 2014.
  [Online]. Available: \url{http://arxiv.org/abs/1411.3406}
\BIBentrySTDinterwordspacing

\bibitem{BT09}
A.~Beck and M.~Teboulle, ``A fast iterative shrinkage-thresholding algorithm
  for linear inverse problems,'' \emph{SIAM J. Imag. Sci.}, vol.~2, no.~1, pp.
  183--202, Jan. 2009.

\bibitem{hubara2016bnns}
I.~Hubara, M.~Courbariaux, D.~Soudry, R.~El-Yaniv, and Y.~Bengio, ``Binarized
  neural networks,'' in \emph{Adv. Neural Inf. Process. Syst.}, Dec. 2016, pp.
  4107--4115.

\bibitem{jaeckel2014quadriga}
S.~Jaeckel, L.~Raschkowski, K.~Borner, and L.~Thiele, ``{QuaDRiGa}: A 3-{D}
  multi-cell channel model with time evolution for enabling virtual field
  trials,'' \emph{IEEE Trans. Antennas Propag.}, vol.~62, no.~6, pp.
  3242--3256, Jun. 2014.

\bibitem{chollet2015keras}
F.~Chollet \emph{et~al.}, ``Keras,'' \url{https://keras.io}, 2015.

\bibitem{tensorflow2015-whitepaper}
M.~Abadi \emph{et~al.}, ``{TensorFlow}: Large-scale machine learning on
  heterogeneous systems,'' 2015, software available from tensorflow.org.

\bibitem{kingma2015adam}
D.~P. Kingma and J.~Ba, ``Adam: A method for stochastic optimization,'' in
  \emph{Proc. Intl. Conf. Learn. Representations}, May 2015.

\end{thebibliography}


\balance

\end{document}